# An image segmentation algorithm based on multi-scale feature pyramid network

Yu Xiao[1], Xin Yang[2], Sijuan Huang[2], Lihua Guo[1]

1. School of Electronic and Information Engineering, South China University of Technology, Guangzhou, Guangdong Province 510641, China.
2. Department of Radiation, Sun Yat-sen University Cancer Center; State Key Laboratory of Oncology in South China; Collaborative Innovation Center for Cancer Medicine; Guangdong Key Laboratory of Nasopharyngeal Carcinoma Diagnosis and Therapy, Guangzhou, Guangdong, 510060, China

Medical image segmentation is particularly critical as a prerequisite for relevant quantitative analysis in the treatment of clinical diseases. For example, in clinical cervical cancer radiotherapy, after acquiring subabdominal MRI images, a fast and accurate image segmentation of organs and tumors in MRI images can optimize the clinical radiotherapy process, whereas traditional approaches use manual annotation by specialist doctors, which is time-consuming and laborious, therefore, automatic organ segmentation of subabdominal MRI images is a valuable research topic.

In the field of automatic segmentation in medical image, U-Net, proposed by Ronneberger et al. [1] in 2015, still has an irreplaceable influence today. Many transformers of U-Net network are proposed, and various plug-and-play components use it as a backbone network [3-10]. Image semantic segmentation differs from image classification. The output of classification is merely a kind of probability without requirements of recovering spatial structure information; whereas segmentation needs coding and decoding of both spatial and semantic information. For reducing the memory acquirement and expanding receptive field, there are pooling layers in segmentation methods. Pooling operations always cause progressive loss of spatial relationships. The U-Net structure is an ingenious design that it allows the features of coding layer to be reused during feature decoding using jump connections. It seems that the loss of spatial relationships can be perfectly solved in U-Net, however, there are still some problems as following, 1) Although the features of encoding and decoding have the same resolution, and can be stitched together, the features of decoding have some information loss comparing with those of encoding. Therefore, they can exist the semantic gap. 2) After multiple convolutional and pooling operations, the spatial location and distribution of features have misalignment problems. Particularly, in the segmentation of subabdominal MRI images, the anal canal and rectal are relatively small size with irregular structures. If spatial relationships of organs can be fully utilized and the semantic gap of encoding and decoding can be solved, the segmentation performance will be improved. Therefore, two solutions are proposed in this paper, which are feature extraction by dilated convolution and multi-scale

This work was supported by Guangdong Basic and Applied Basic Research Foundation(2022A1515011549) and
(Yu Xiao, Xin Yang and Sijuan Huang have made equal contributions to this work.) (Corresponding author: Lihua Guo, Email: guolihua@scut.edu.cn.)

feature pyramid network. The main purpose of them is to extract multi-scale information, and to maintain spatial information and semantic uniformity between encoding and decoding.

## 1. The related research

The essence of image segmentation can be seen as the classification of every pixels. Long [2] et al. proposed a fully convolutional network (FCN) in 2015, which enables end-to-end prediction of image segmentation. In addition, Ronneberger et al. [1] proposed U-Net to improve FCN. U-Net consists of three parts, the encoding, the decoding and a jump connection between them. Because of the good performance, its network structure has been widely used in medical image segmentation. From them on, many network structures based on U-Net have been proposed, such as attention mechanisms, densely connected module, cascade network, etc.

Biomedical images differ from natural images in that they are a set of 3D image slices at different locations. The spacing between slices usually varies from 2 to 3 mm, so using traditional 2D models to process 3D medical images has the following problems, Firstly, it is inefficient because 3D data is treated as series of 2D data. Secondly, valuable spatial information is ignored. In order to make better use of spatial information, Çiçek et al. [3] proposed a 3D U-Net, which takes 3D images as input and output, improving the segmentation accuracy of U-Net model in 3D images. The V-net proposed by Fausto Millemari et al. [4] has similar structure to 3D U-Net, and a residual structure is added to avoid gradient disappearance. The authors use convolution with a step size of 2 to take place of pooling. When the input image is a 3D image, the corresponding convolution method also changes to 3D convolution, which causes exponential increasement in computational costs. DeepMedic proposed by Kamnitsas et al. [6] adopts a compromise approach by selecting properly-sized image block and using full convolution operation to make dense predictions for multiple neighboring pixel, which reducing computational complexity. Due to the regularity in medical images, the selection of image blocks can solve problems of intra-class data imbalance under the control of the scale of the foreground background. Ibtehaz et al. [7] proposed a MultiRes-UNet method that includes residual path, which allows the encoder features to perform some additional convolution operations before being fused with the corresponding features in the decoder. These convolution operations are used to balance the depth gap between the sibling encoder and the decoder. Seo et al. [8] argued that traditional jump-join operation was too brutal. For large targets, the dual operations of jump-connection and encoding-decoding are repetitive for low-resolution information, and it will blur the boundaries of segmented target. For small targets, direct transfer cannot effectively learn enough high-resolution features. They proposed the mUNet, which adaptively merges features in residual path into the jump connection. The FED Net proposed by Chen et al. [9] thinks that features of different resolutions have semantic differences, and designs feature fusion structures to integrate these features into the encoder. Both mUnet and FEDnet incorporate convolutional operations in the jump connection to improve the performance of segmentation in medical images. Oktay et al. [10] incorporated an attention gating system at the end of the jump connection to optimize segmentation by using features of next level to supervise features of the previous level, limiting the activation to the region to be segmented and reducing the activation value of the background.

## 2. Feature reuse based on dilated convolution

The U-Net network structure is divided into two parts: encoding and decoding. Take the 2D U-Net network as an example, its input is a 1-channel CT image. The encoding path consists of four encoding levels. The first encoding operation changes the input channel from 1 to 32, with a convolutional kernel size of 3 x 3 and a step size of 1. The subsequent three-stage coding consists of three steps: convolution, Batch Normalization, and the nonlinear activation function (RELU) repeated twice respectively. Each encoding operation is preceded by a maximum pooling to down-sample the features. The decoding path consists of four levels of decoding, having the same network structure as encoding. Each level of decoding is tightly followed by an up-sampling operation, which is done by nearest-neighbor linear interpolation. The output of decoding turns the number of channels into the total number of organ image channels in final segmentation, and the pixel value of each channel represents the probability that the pixel belongs to a particular organ (or background) under that channel.

the proposed feature reuse based on dilated convolution is mainly implemented as follows, assuming $X_E$ and $X_D$ represent the feature mapping in encoding (Encoder) and decoding (Decoder) respectively, $X_F$ denoting the feature mapping of intermediate process. Given input $I \in R^{1 \times H \times W}$, the convolutional neural network obtains features $X_E^i$ at different scales by encoding and pooling for $i \in (1,2,\cdots n)$ times, where $X_E^i \in R^{C_i \times \frac{H}{2^{i-1}} \times \frac{W}{2^{i-1}}}$. For each $X_E^i$, the following operations are performed:

$$X_F^i = D_r\big(H(X_E^{i-1})\big)$$

where $D_r(\cdot)$ denotes a dilated convolution with a dilation rate of $r$ and $H(\cdot)$ denotes a single encoding operation containing convolution, batch normalization and non-linearization process. $X_F^i$, the output of this process, is a new feature having the same resolution as $X_E^{i-1}$, $X_F^i \in R^{\frac{C_i}{2} \times \frac{H}{r} \times \frac{W}{r}}$. At the end of $i$ encoding, the input $X_D^1 = [X_F^i]$ to the decoding network is obtained, where $[\cdot]$ represents the splicing operation. The decoding operation can be expressed as follows,

$$X_D^j = U\big(H(X_D^{i-1})\big) + X_E^i$$

Where $U(\cdot)$ denotes up-sampling, $X_E^i$ are features from encoding network, having the same number of channels and resolution as $U\left(H(X_D^{i-1})\right)$.

Unlike traditional methods, the input to decoding layer in this network is not the output of the last layer of encoding stage, but the same size features obtained by dilated convolution from encoded information at different scales with different expansion rates, and they are stitched together along the channel dimension as the input to the decoding layer. This operation not only allows the intermediate results of the encoding layer to be reused, but also allows the high-resolution spatial location relations to be transmitted more directly to the low-resolution lawyer, enhancing the spatial representations during decoding, especially for features in early stages of decoding. In jump connection, the algorithm replaces the splicing operation with a point-by-point sum operation, which reduces the number of parameters, and the residual structure effectively avoids the problem of gradient disappearance. The most important reason is that the features of proposed network generated during encoding have already been partially extracted by the structure of feature multiplexing. To avoid repeated superposition, we chose point-by-point sum operation instead of splicing operation.

## 3. Multi-scale Feature Pyramid Network

In order to achieve good spatial location feature reuse in the information decoding process, multiple spatial reuse structures are used to form a feature pyramid network, realizing more sufficient feature reuse and more complete spatial information recovery strategy. Therefore, a multi-scale feature pyramid network (MFP-Net) is added to the network structure and its implementation are as follows: assuming given input $I \in R^{1\times H\times W}$, the convolutional neural network obtained features of different scales by $i$ encoding, $X_D^i \in R^{C_i\times\frac{H}{2^{i-1}}\times\frac{W}{2^{i-1}}}$. For each encoded feature at different scale, the following operations are performed:

$$X_F^{ik} = D_{r(k)}\left(H(X_E^{i-1})\right) k \leq n-i$$

where $D_{r(k)}(\cdot)$ denotes the dilated convolution with a dilation rate of $r(k)$ and $n$ is the total number of encoding. $X_F^{ik}$ is a set of new features at different scales, $X_F^{ik} = D_{r(k)}(X_i) \in R^{\frac{C_i}{2}\times\frac{H}{r(k)}\times\frac{W}{r(k)}}$. The input to decoding network is $X_D^1 = \left[X_F^{i1}\right]$. The decoding operation can be expressed as:

$$X_D^j = U\left(H(X_D^{j-1})\right) + X_F^{ik}$$

The repeated use of this module is very helpful for the gradual recovery of spatial information. In the pyramid structure, not only the different scale features undergo different expansion rates for scale unified, but also the same scale features undergo dilated convolution with different expansion rates, and distributed to decoding features at different scales respectively, which provides more possibilities for the recovery of feature information. From another perspective, the multiplexing structure facilitates the fusion of features at different scales in an explicit way, while decoding network chooses effective information that is more useful for the decoding process, semantic reinforcement, and spatial recovery through training, rather than simply splicing with features of the same resolution in self-coding. In addition, the calculation mechanism of the number of channels allows the maximum information of channels from the corresponding resolution of the encoding layer to be obtained at the current resolution, in order to maintain the maximum utilization of information at the same resolution. The network structure where such coded features are multiplexed through dilated convolution is called Multi-Scale Feature Pyramid Network (MFP-Net), whose schematic diagram of structure is shown in Figure 1. In the middle of the network, i.e. the pyramid structure, encoding features of different scales are multiplexed through various dilated convolutions with different expansion rates, and features of the same scale are stitched together, for which more feature multiplexing modules are merged in decoding process, thus gradually enhancing the spatial constraints of decoding process.

To explain the mechanism of this method, this paper obtains the characteristic heat map during training process, as shown in Fig. 2. Because the spatial position alignment is completed by applying dilated convolution, the four feature maps can have the same physical resolution, where (a), (b) and (c) are results obtained from dilated convolution of features at different scales with different expansion rates respectively. (d) are the decoded features generated by corresponding decoding network (for presentation purposes, only one of the feature maps has been chosen as representative). It can be clearly observed that the features contained in images are becoming more and more abstract (advanced), while the spatial information of the image is becoming more and more blurred. In this paper, in order to allow the decoding feature (d) to recover the space appropriately in advance, the more spatially constrained (a) and (b), as well as the equally abstract feature (c), are jointly passed into (d) , with completing the extraction of spatially critical information and the re-fusion and enhancement of semantics in one decoding process.

## 4. Experiments and analysis of results

### 4.1 Pre-processing of experimental data

In order to facilitate the subsequent label visualization, the Nibabel library in python was used to parse the raw data (medical DICOM format) and extract its image data as well as the corresponding label data. Meanwhile, five main organs of the pelvic region were selected for segmentation: anal canal, bladder, rectum, femoral head (left) and femoral head (right).

The data were firstly processed as follows to ensure fairness and robustness of the experiment:

(1) Some of the data had different scales in the three sequences, i.e. the slices of three sequences showed different ranges of images shot in the same position from the same patient, requiring rough alignment of the image data and corresponding labels.

(2) The original images were of various sizes such as 768 x 768 and 560 x 560. To ensure uniformity of data, taking model size, memory consumption and the position of external contours in images into account, all images in this paper were scaled down to 384×384 and then cropped to 256 (height) × 384 (width). It should be noted that in final test, the labels are filled to 384×384 and scaled up to their original size and compared to the original labels.

(3) The training sample of experiment is only 55 cases with 3245 slices, which is not enough for training compared to common image dataset of tens of thousands of images. The data enhancement techniques are used to make the model more robust by using random cropping and small rotations, etc.

(4) The image segmentation labels of the dataset are in pixel points, with six numbers from 0-5 representing different parts respectively, as shown in Table 1.

## 4.2 Experimental parameter settings

The computing device is an RTX3090 with 24GB of video memory size, the programming language is python 3.8, and the experimental model is deployed on PyTorch machine learning framework in PyTorch version number 1.9.0, with parallel computing on CUDA platform in version 11.1 and using NVIDIA cuDNN (CUDA Deep Neural Network) to accelerate the network.

Other training parameters for the network were as follows: the data were trained in batches with batch size of 32, for a total of 400 generations. The initial learning rate was set to 0.01 and decreased to 0.001 and 0.0001 at the 200th and 300th generations respectively. SGD was used as optimizer for the network, setting the momentum term to 0.9 and the weight decay factor to 0.001.

Under different network models, additional different parameters were set to fit the network without affecting the overall parameters (e.g. in the R2UNet comparison experiments, truncated gradients were set to avoid gradient explosion). In this paper, weighted cross-entropy loss is used as the loss function for training, with a weighting ratio of 1:10:4:10:8:8 for the loss function.

The experiments were conducted by randomly dividing the 55 cases of training data into 5 groups of about 650 slices each. The model that performed best on validation set was selected by 5-fold crossover experiment, and then tests on the remaining 28 cases that did not involve in training process were conducted.

## 4.3 Evaluation metrics

Two evaluation metrics were used to evaluate the performance of model, namely the three-dimensional Dice Coefficient ( （Volumetric Dice Coefficient, DSC) and the Mean Surface Distance (Mean Surface Distance, MSD). The Dice coefficient is used to measure the degree of overlap between two samples and takes values in the range [0,1].The larger the Dice coefficient , the more similar the two samples are. The mean surface distance (MSD) is used to measure the physical distance between the samples, and the smaller the MSD, the closer the samples are to each other. The formula calculating DSC is as follows.

$$\mathrm{DSC} = \frac{2TP}{FP + 2TP + FN}$$

where TP is true positive, FP is false positive, FN is false negative, and TN is true negative. They indicate that the pixel is correctly classified, misclassified, underclassified, and the background pixel is correctly classified respectively. MSD is calculated as follows,

$$MSD(A,B) = \frac{1}{|S(A) + S(B)|}\left(\sum_{a\in S(A)} d\big(s_A, S(B)\big) + \sum_{b\in S(B)} d\big(s_B, S(A)\big)\right)$$

where $S(A)$ is the set of surface points in $A$ and $S(B)$ is the set of surface points in $B$ , $d\big(v,S(A)\big)=\min_{s_A\in S(A)}\|v-s_A\|$. That is, for all surface points of one of the objects, the minimum value of Euclidean distance from this point to all points of another object is calculated and finally averaged by accumulation.

## 4.4 Analysis of experimental results

For selection of the comparison algorithm, the most influential U-Net structure was chosen firstly. As the point-by-point addition is used instead of splicing operation in MPF structure, the splicing operation of jump connection in U-Net was directly modified to the form of point-by-point addition (U-Net-Add) to demonstrate the effectiveness of MFP. In the selection of other methods, more attentions were paid to adaptation of the network form, such as the UNet++ structure with variable depth properties, r2u-net which introduces recurrent neural networks into segmentation, and MedT which uses the currently visually popular Transformer structure. To indicate distinctions, the method that used only dilated convolution multiplexing structure was named MFP-Net1, while the method using both dilated convolution and multiscale feature pyramid structures was named MFP-Net2.

Apart from visual comparison in experimental results, the number of parameters and model size were added to more fairly measure the differences to results causing by different methods. The first convolution operation of the U-Net was used to generate 8 channels, 16 channels, and 32 channels (the number of channels increases exponentially accordingly), representing different U-Net structures with different number of parameters.

In order to compare methods from different dimensions, two currently widely-accepted evaluation metrics, the Dice coefficient and the mean surface distance, were selected. The experimental data under the two metrics are shown in Tables 2 and 3 respectively.

Comparing U-Net results of three different channels in Table 2, it is found that a greater number of parameters does not bring more effect gain, and a smaller number of parameters will also reduce prediction accuracy due to insufficient fit. Thus, the algorithm in this paper choose the U-Net16 structure, which is relatively balanced in terms of effect and parameters, as the backbone of network structure, and modifies it based on U-Net16. As shown in Table 2, MFP-Net1 and MFP-Net2 not only have fewer parameters than U-Net16, but also have significantly better prediction results than U-Net16. The spatial multiplex structure based on dilated convolution (MFP-Net1) achieves best Dice coefficient values in the anal canal and femoral head regions(right), with Dice coefficients 0.031 and 0.008 higher than that of U-Net respectively. The pyramidal network based on multiplexed structure (MFP-Net2) achieved best Dice coefficient values in the bladder and femoral head regions(left), with Dice coefficients 0.017 and 0.021 higher than that of U-Net respectively.

As seen in Table 2, the MFP-Net has fewer parameters compared to U-Net16 in both number of parameters and model size, due to the point-by-point summation used in the feature multiplexing structure instead of splicing. The reason for less parameters in pyramid network is the full use of multiplexing structure, which uses channels from each encoding layer and halves the number of channels in encoding layer, thus greatly reducing the number of channels in decoding layer.

It is worth noting that MedT, which applies the Transformer structure to medical image segmentation, has very little effect on this task, instead bringing more than five times the training time, even slower fitting speed, and of course greater memory requirements. Its performance on Dice coefficient is the worst, but its performance on MSD is not too bad. The reason is that the Transformer structure used by MedT has a strong ability to capture contextual remote dependencies, and MSD is particularly sensitive to outlier sample points.

Specifically, for each case, box-line plots were used to show the distribution characteristic of statistical data, as shown in Figures 3-5, from which it can be seen that the traditional U-Net approach performs worse for those difficult samples.

Based on above analysis, a basic conclusion can be drawn: the feature multiplexing structure based on dilated convolution and the resulting extended multi-scale feature pyramid network MFP-Net can effectively improve the segmentation performance of organs while using fewer resources. In order to present and compare the segmentation results more clearly, the segmentation results of selected labels and some representative methods were visualized separately in three dimensions from different perspectives, as shown in Figure 4, with different colors representing different organs. It can also be seen from the figure that in the first three methods, there are more or less isolated prediction points that deviate from the organs, causing increase in false positive values and thus affecting their Dice coefficients. Meanwhile, the more distant the isolated points are, the larger the MSD will be, leading to conclusion that multiscale feature pyramid network can improve the segmentation performance of organ, consistent with the findings in Table 2, Table 3, and Figure 3. MedT, although insensitive to training boundaries, is also good at determining whether a organ is present in the slice based on the overall image information, thus avoiding the generation of false positive results, so its deviation points are yet relatively few (it should be noted particularly that due to limitation of video memory, the experiment reduced the size of input images during MedT training, so the jaggedness of segmentation are stronger in MedT's visual images), but this learning framework had to be abandoned due to the huge video memory requirements of Transformer structure.

## 5.Conclusions

Based on the analysis of shortcomings in traditional encoder-decoder network structure, i.e. jump connections are not able to convey strong constraints on the recovery of spatial information, especially during decoding. In this paper, two network structures are proposed for improvement: feature multiplexing structure based on dilated convolution and multiscale feature pyramid network. The feature reuse structure utilizes the large-scale receptive field of dilated convolution to sequentially extract new features in feature maps generated by encoding, and fuse these new features to obtain features with stronger spatial information than the underlying encoding layer, and they are used as input to the decoding network. The multi-scale feature pyramid network reuses the feature reuse structure for multiple times, fusing corresponding reused features before each decoding, which maximizes constraints on decoding network because the reuse features are derived from pre-coding period and they have strong spatial features. Experiments show that the proposed network structure is able to significantly reduce the false positive problem and has fewer model parameters than U-Net.